\documentclass[journal]{IEEEtran}

\ifCLASSINFOpdf
  \usepackage[pdftex]{graphicx}
 
\else
 
\fi

\ifCLASSOPTIONcompsoc
  \usepackage[caption=false,font=normalsize,labelfont=sf,textfont=sf]{subfig}
\else
  \usepackage[caption=false,font=footnotesize]{subfig}
\fi

\usepackage{url}

\usepackage[latin1]{inputenc}

\begin{document}

\title{ PETALO read-out: A novel approach for data acquisition systems in PET applications}

\author{V.~Herrero-Bosch, R.~Gadea, R.J.~Aliaga, J.~Rodr\'iguez, J.F.~Toledo,
		R.~Torres-Curado, F.~Ballester, R.~Esteve, J.J.~G\'omez-Cadenas and P. Ferrario% <-this % stops a space
\thanks{V. Herrero-Bosch, R. Gadea, R.J. Aliaga, J.F. Toledo, R. Torres-Curado, F. Ballester and R. Esteve are with Instituto  de  Instrumentaci\'on  para  Imagen  Molecular  (I3M), Universitat  Politecnica  de  Valencia, Camino  de  Vera  s/n,  46022  Valencia,  Spain, . e-mail: viherbos@eln.upv.es.}% <-this % stops a space
\thanks{J. Rodr\'iguez is with Instituto  de  F\'isica  Corpuscular (IFIC), CSIC \& Universitat  de  Valencia. Calle  Catedr\'atico  Jos\'e  Beltr\'an,  2,  46980  Paterna,  Valencia,  Spain}% <-this % stops a space
\thanks{P. Ferrario and J.J. G\'omez-Cadenas are with Donostia  International  Physics  Center  (DIPC), Paseo  Manuel  Lardizabal  4,  20018  Donostia-San  Sebastian}% <-this % stops a space
\thanks{This work was supported by Spanish Ministry of Economy and Competitiveness under project FPA2016-78595-C3-3-R and the European Research Council (ERC) under Starting Grant 757829-PETALO.}}


\maketitle

\pagenumbering{gobble}

\IEEEpeerreviewmaketitle

\section{Introduction}

\bstctlcite{viherbos:BSTcontrol}

\IEEEPARstart{N}{owadays} most of the detectors used in positron emission tomography (PET) applications are based on pixelated detectors. Usually a high number of individual scintillator crystals are either coupled individually to a light sensor or sharing a fewer number of them. In both cases the spatial resolution is limited by the pixel size. The high degree of segmentation in those detectors strongly simplifies the task of locating gamma ray interaction although on the other hand small pixel size degrades energy resolution and time performance.  
Moreover the scalability of pixel based architecture requires a high cost in terms of material and manufacturing. Other solutions have been proposed in the past \cite{HERRERO11} based on bigger scintillation crystals where the light distribution sensed by an array of photodetectors was analyzed in order to obtain the gamma event characteristics. However the size and shape of the scintillator crystal cannot be fully customized and large area detectors (body PET scanners) are still difficult to build.

\section{PETALO detector concept}

\par
\begin{figure}[!t]
	\centering
	\includegraphics[width=3.5in]{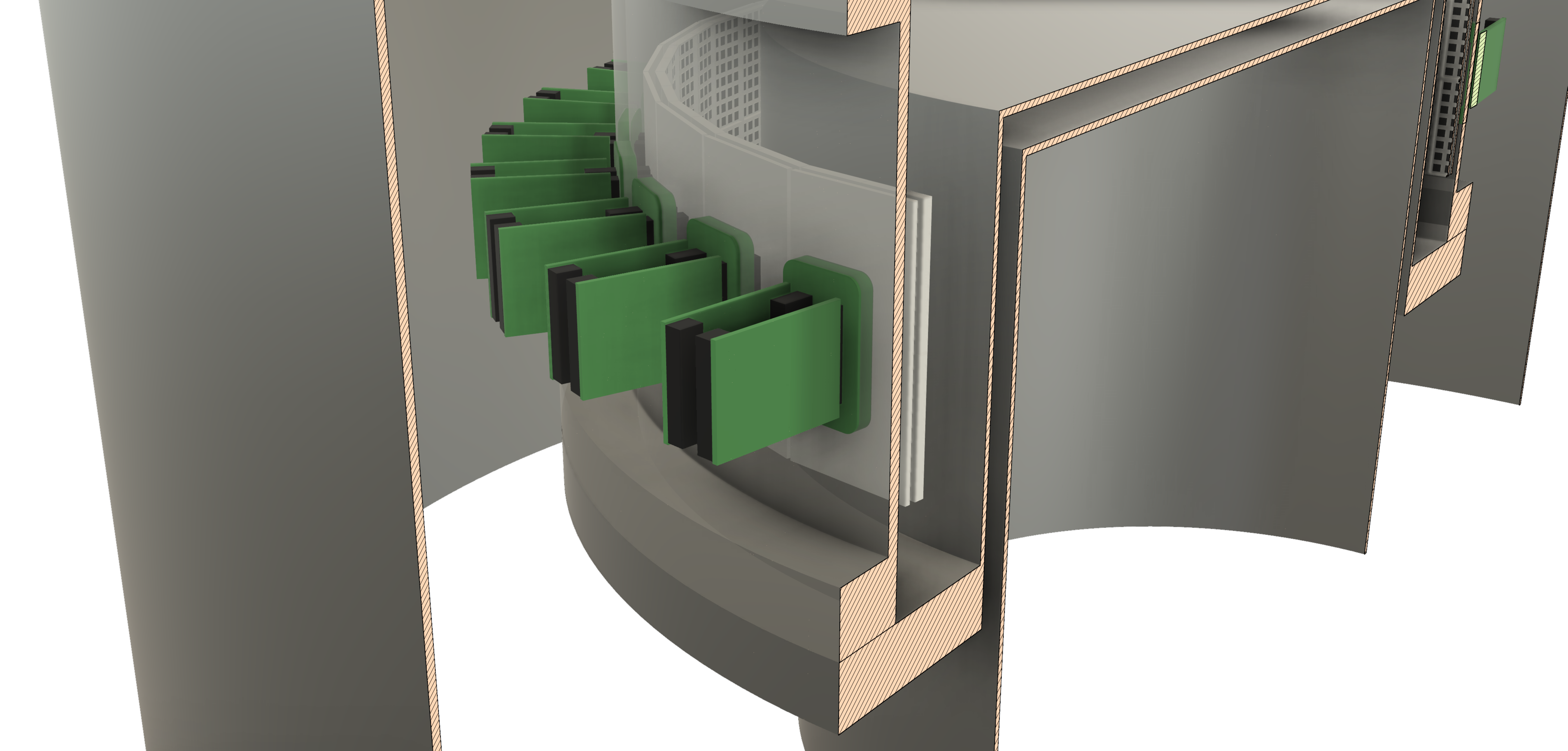}

	\caption{Cryostat for PETALO prototype.}
	\label{cryostat}
\end{figure}

 PETALO (Positron Emission TOF Apparatus based on Liquid xenOn) detector makes use of liquid Xenon (LXe) as scintillation medium which not only offers interesting properties such as high scintillation yield ($\sim$30k photons per 511 keV gama) and fast time performance \cite{GOMEZ_CADENAS16} but also allows to build custom size detectors suitable for large volume applications. Since LXe requires a cryogenic environment (cryostat) some parts of the detector must be compatible with low temperature operation. However most of the sensitive electronics can be placed in the vacuum space of the cryostat device thus avoiding the effects of cryogenic temperatures (see fig.\ref{cryostat}).

\par
\begin{figure}[!t]
	\centering
	\includegraphics[width=3.5in]{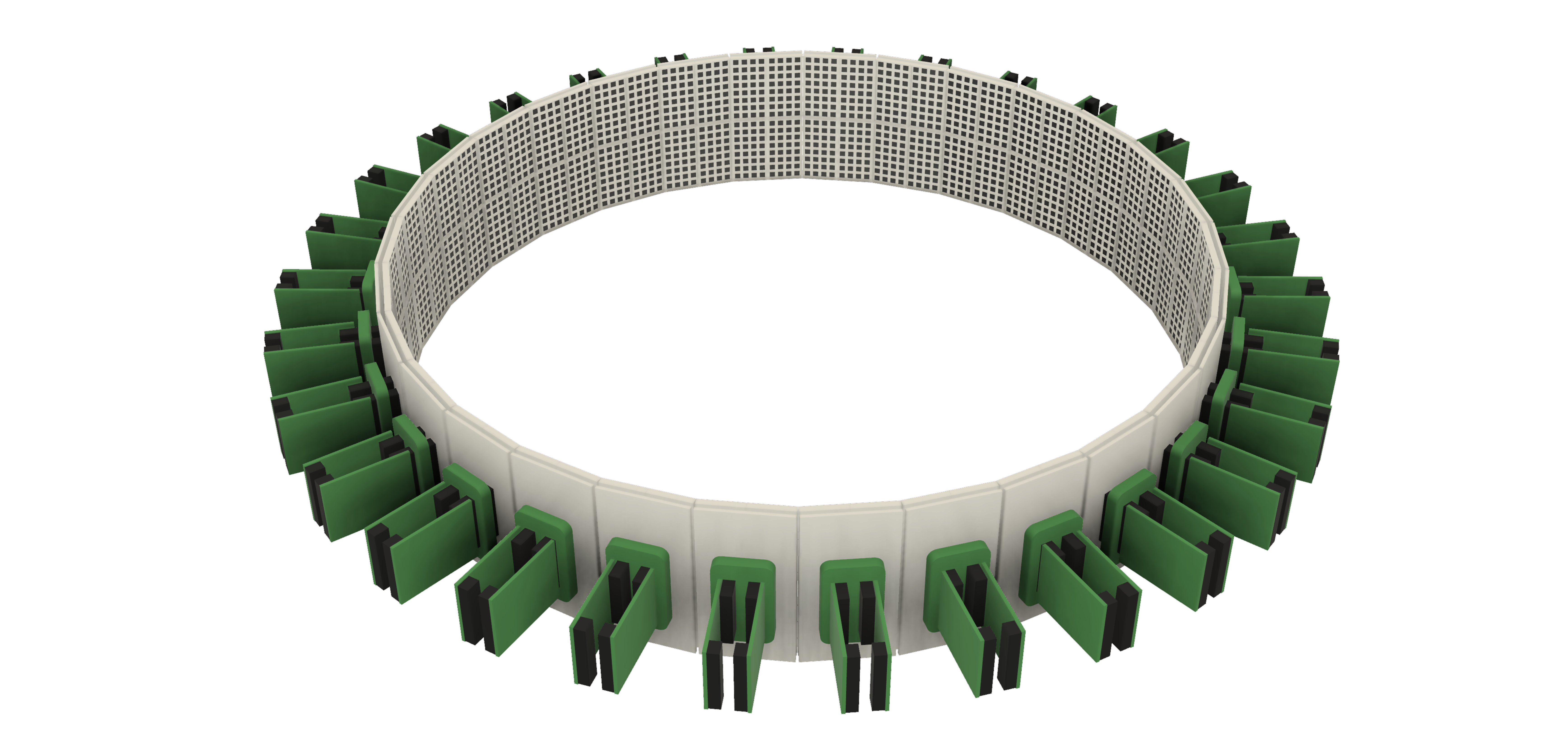}
	\caption{PETIT prototype detector structure.}
	\label{PETIT}
\end{figure}

 PETALO detector aims at capturing the light produced by the scintillation in LXe taking advantage of its uniform response and continuity. This strategy will lead to a geometrical distortion free behavior compared to other PET detectors. In order to achieve the best characteristics, densely packed arrays of VUV sensitive SiPMs will be employed to meet LXe scintillation light wavelength requirements. The enhanced light collection capabilities of such structure will increase resolution performance of previous LXe based detectors \cite{WASEDA04}. Moreover the low temperatures inside the cryostat will make dark count rate effects neglectable. PETIT prototype, the first version of the PETALO detector, which is being built at the moment will be composed of a 17 cm height and 36 cm diameter ring with a fully instrumented outer face (fig. \ref{PETIT}). In later versions the height of the ring will be extended in order to further reduce border effects in the field of view (FOV).

\par In order to take advantage of the unique PETALO detector characteristics a read-out architecture must be designed to meet the following specifications: Electronics associated to detector (front-end and read-out itself) must be fully expandable in terms of detector size. Read-out scheme must be compatible with the non-segmented structure of the detector. Time of Flight (TOF) capabilities must be assured at the scanner level, that is to say front-end and read-out electronics should not degrade LXe time performance \cite{GOMEZ_CADENAS16}.

\subsection{Brain PET case study}
\IEEEpubidadjcol
\par Continuous systems show a high spatial pileup probability as well as a much higher amount of information generated for an equivalent activity in the FOV. Light distribution for every gamma detected must be acquired and sent for further processing and coincidence analysis. PETALO design allows to extract more information from every event and eventually will implement Compton reconstruction algorithms in order to enhance overall detector sensitivity. However continuous medium detectors are more prone to pile up effects since more individual sensors get involved in every event. A typical brain PET application can be used as an example to show a worst case pile-up probability as well as the total amount of information generated by the system. For the usual dose of FDG in such a PET scanner assuming the dimensions of PETIT prototype the pile up probability is shown in table \ref{activity}. On the other hand an approximate computation of the information generated is presented in table \ref{information} for different implementation options. Although estimations show the feasibility of PETALO concept, the total data flow required through the readout system is high enough to require a new design approach in order to optimize the data link speeds as well as introduce data reduction or compression techniques in its first stages.

\begin{table}[!t] 
\renewcommand{\arraystretch}{1.3}
\caption{Brain PET activity and pile-up}
\centering
\label{activity}
\begin{tabular}{|c|c|}

	\hline 
	TOTAL FDG dose & 240 [MBq] (peak) \\ 
	\hline 
	Brain activity & 36 [MBq] (peak) \\ 
	\hline 
	Detected Activity & \\
	(100\% efficiency) & 30.7E6 [$\gamma$/s] (peak) \\ 
	\hline 
	Pile-up probability & 10.4 \% \\ 
	\hline 
	
\end{tabular}  
\end{table}

\begin{table}[!t]
	\renewcommand{\arraystretch}{1.3}
	\caption{Brain PET information generation. (63 \% detection efficiency)}
	\centering
	\label{information}	
	\begin{tabular}{|c|c|}

		\hline 
		Bits per channel & 46 \\ 
		\hline 
		Bits per event & 1472 \\ 
		\hline
		Raw data rate (average) & 25.1 Gbit/s \\ 
		\hline 
		Data rate for coincidence filtering & \\
		integrated in readout & 5 Gbit/s \\ 
		\hline 
		Data rate for coincidence & \\
		and energy filtering integrated in readout & 0.7 Gbit/s \\ 
		\hline

	\end{tabular}  
\end{table}

\section{PETALO read-out proposal}

\par
\begin{figure}[!t]
	\centering
	\includegraphics[width=3.5in]{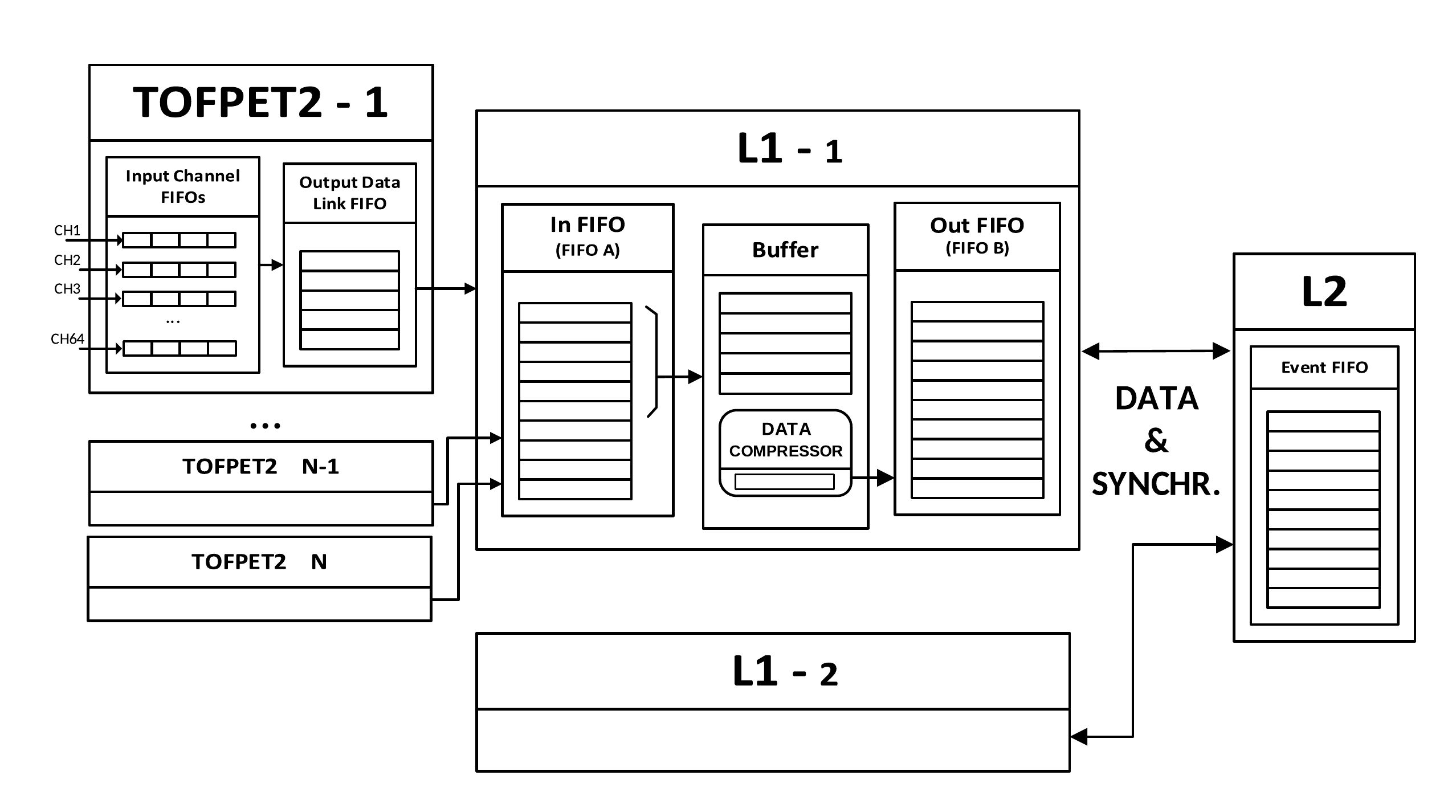}
	% where an .eps filename suffix will be assumed under latex, 
	% and a .pdf suffix will be assumed for pdflatex; or what has been declared
	% via \DeclareGraphicsExtensions.
	\caption{Read-out scheme.}
	\label{read_out}
\end{figure}

 A front-end and read-out architecture has been proposed (fig. \ref{read_out}) to meet the stated requirements. Front-end section (FE) is based on TOFPET2 asics \cite{TOFPET2} which offer a good solution for early digitization of sensor outputs and timestamp generation with a CRT resolution close to that of the LXe. Each TOFPET2 is able to handle 64 SIPMs generating timestamps and integrated charge data for every channel individually. A certain number of TOFPETs are controlled by a first level (L1) readout board that manages their configuration and receives their output data. L1 main function is to reduce the amount of information generated by FE and send it to upper levels of the readout architecture. A second level (L2) readout board will finally group all the channel data with same timestamp and look for gamma event coincidences in order to build true detected events. In an initial stage of development L2 functionality can be carried out by a dedicated computer able to handle the resulting data bandwidth.

\par The output data flow from the detector shows statistical variations due to two dominant effects. The most important is the Poisson fluctuation in the event rate but also the analog to digital conversion  (ADC) done at the FE level has a variable latency due to the usage Wilkinson ADC converters. In order to attenuate the data flow fluctuations a set of FIFO based memory buffers must be introduced between the different stages of the readout. The depth of these FIFOs along with the data link speeds in every point of the readout scheme are fundamental parameters for the design process. A data flow simulation test bench has been developed to get a realistic approximation for these values. The physics data generation is based on a GEANT simulation which uses the same event rate as the one obtained in the brain PET study. The detector architecture chosen for this simulation is a fully SiPM instrumented PETALO prototype with the same geometry specifications as the one used in the previous pile up study. A total number of 7680 SiPMs covering the inner face of the cylindrical volume of the detector are controlled by 120 TOFPET2 asics and 12 L1 (10 asics per L1) in this configuration. Behavioral data flow models with latency and timing specifications have been developed for all the elements and whole parametrized readout architecture has been simulated under different conditions using Python based SIMPY package \cite{SIMPY} to get the required information. 

\par Another key element of the PETALO readout concept is the synchronization scheme required to generate coherent timestamps at the FE level. Every TOFPET asic in the system needs to be synchronized within a 50 ps margin in order not to degrade the timing resolution of captured events. Self-synchronizing links are deployed between L1 and L2 boards according to the same principles described in \cite{ALIAGA14}, that allow propagation of a common clock frequency and automatic correction of timestamps.

%First task implemented in L1 is to store all data received from TOFPET2 channels with same timestamp. Thus every L1 will collect a piece of the light distribution of a detected event.
% which will allow to apply some data compression algorithm to reduce data link bandwidth requirements
%\par A typical PET configuration can generate such a huge amount of data that data reduction mechanism is required along with the L1 lossless compression. Two different energy thresholds (TE$_A$ and TE$_{L1}$) have been introduced in order to block all the channels under a certain charge level. TE$_A$ is applied at TOFPET2 level and reduces data rate between TOFPET2 and its L1. This first threshold has a destructive behavior for global energy computation since information related to channels below threshold will be permanently lost. The second threshold TE$_{L1}$ is implemented in L1 and reduces the data rate at its output. However all the equivalent charge in the channels below this threshold is added together and sent as a complementary field inside the compressed data frames generated by L1. As a consequence TE$_{L1}$ effect on total energy computation is non-destructive.

%\begin{figure*}[!t]
%\centering
%\subfloat[Reference Configuration]{\includegraphics[width=3.6in]{./figures/daq_GOLDEN}%
%\label{simulation_ref}}
%\hfil
%\subfloat[Parameter effects in L1]{\includegraphics[width=3.5in]{./figures/trends.pdf}%
%\label{simulation}}
%\caption{Read-out simulation results}
%\label{simulation_ab}
%\end{figure*}

\section{Read-out system design and simulation}

\par In this section a description of the functionality and simulation results of every component in the readout architecture will be presented. The simulation results are based in the brain PET study case shown in the previous section.

\subsection{Front end modelization}
\par
\begin{figure}[!t]
	\centering
	\includegraphics[width=3.25in]{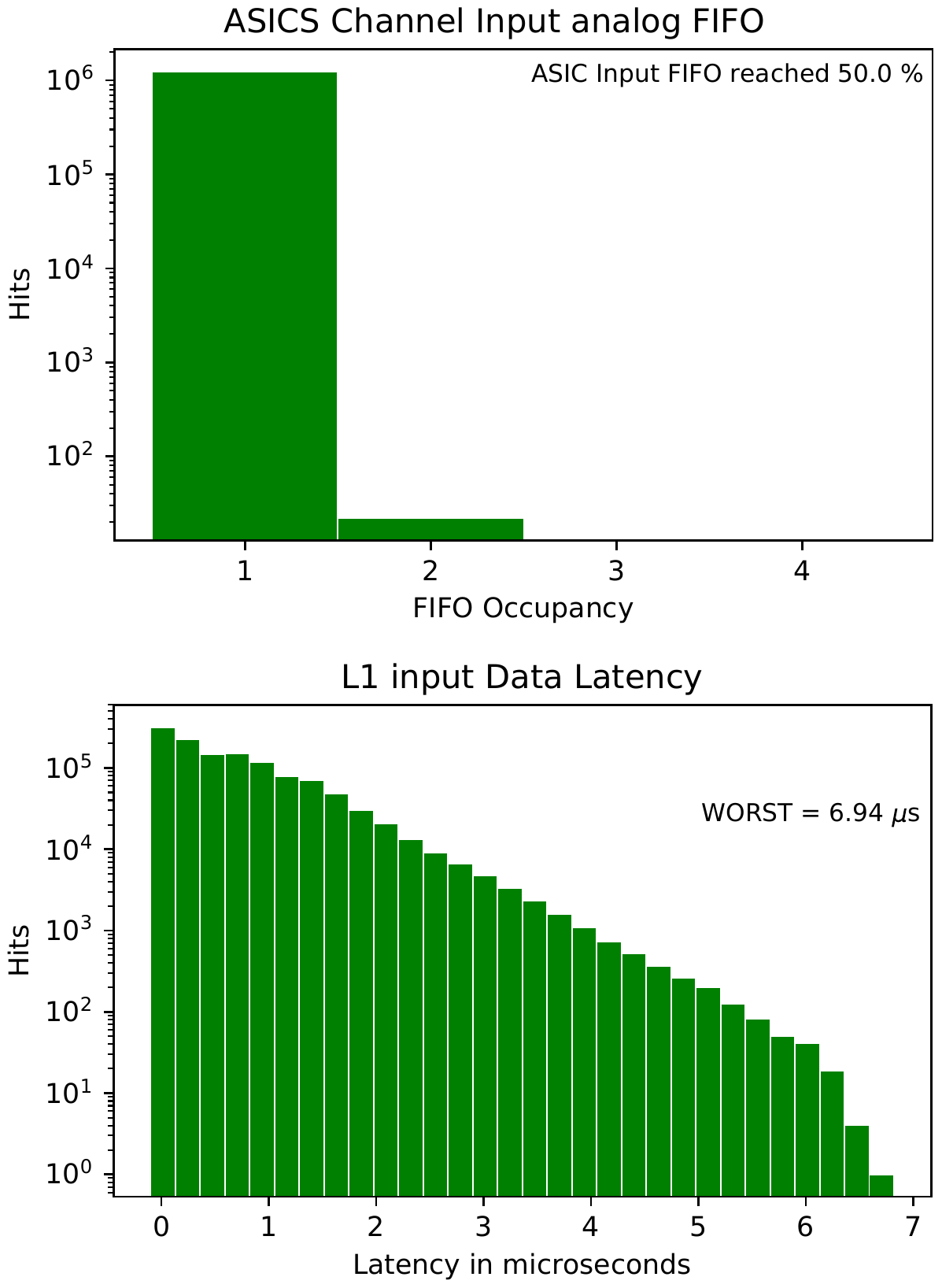}
	% where an .eps filename suffix will be assumed under latex, 
	% and a .pdf suffix will be assumed for pdflatex; or what has been declared
	% via \DeclareGraphicsExtensions.
	\caption{FE simulation data.}
	\label{FE_sim}
\end{figure}

 A behavioral model for TOFPET asic has been developed based in its internal structure. Each input channel has a 4 depth analog FIFO, a Wilkinson ADC converter with a maximum latency of 5 microseconds and a time to digital converter (TDC) for timestamp generation. The output digital link of the device has been modeled with a speed of 32 Mevents/s which is lower than the maximum provided by the manufacturer specifications. Results for input analog FIFOs and maximum data latency in the FE stage are shown in figure \ref{FE_sim} for a 50K events simulation.

\subsection{L1 board modelization}
\par
\begin{figure}[!t]
	\centering
	\includegraphics[width=3.25in]{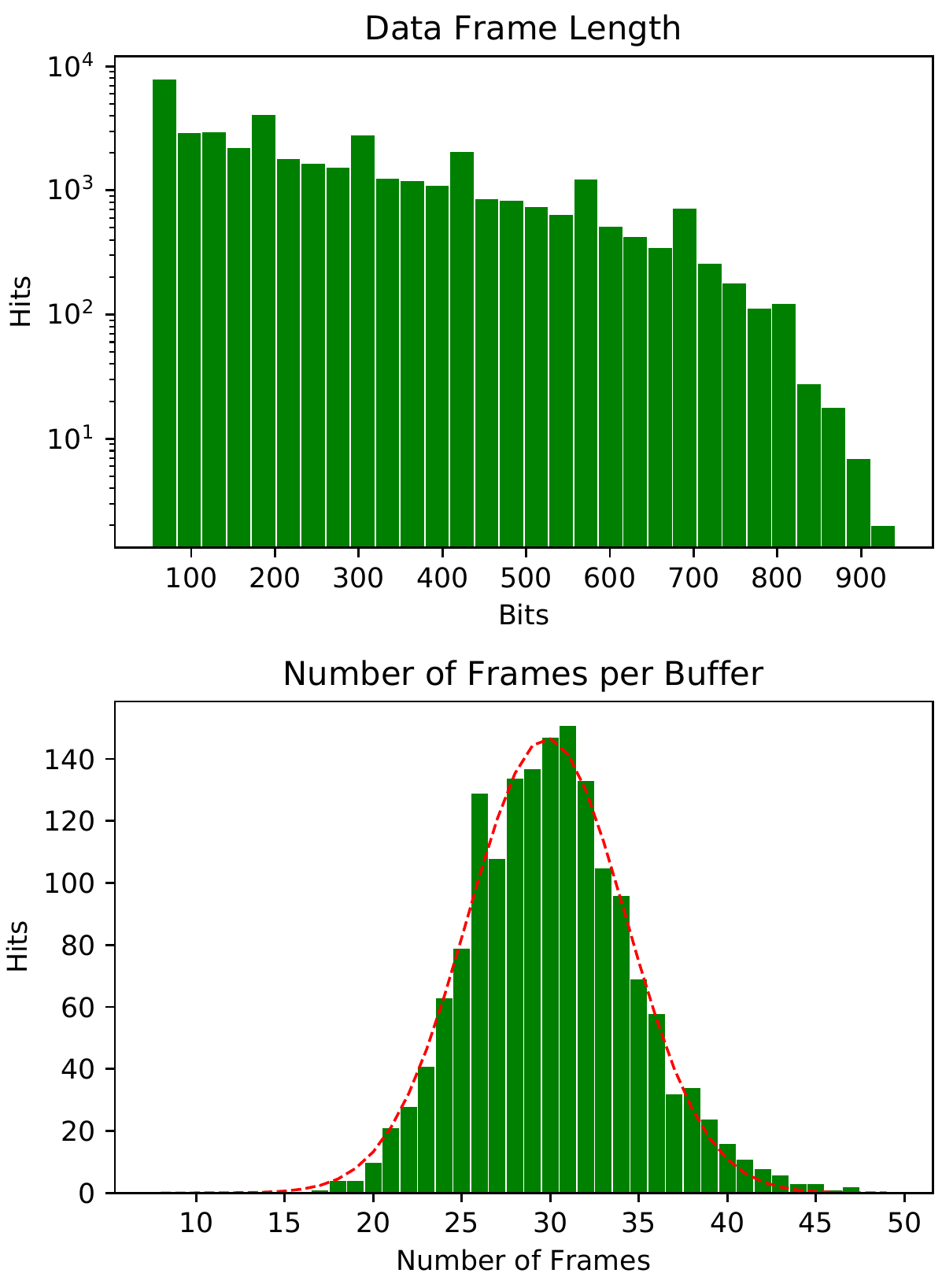}
	% where an .eps filename suffix will be assumed under latex, 
	% and a .pdf suffix will be assumed for pdflatex; or what has been declared
	% via \DeclareGraphicsExtensions.
	\caption{L1 simulation data.}
	\label{L1_sim}
\end{figure}

 First level readout boards (L1) must collect the information of FE stage and compose the pieces of the light distribution associated to each gamma event detected. Since each channel generates data with different arrival times to the L1 input, each L1 board must store a batch of data in a buffer that will have to be scanned continuously to build data frames with the same timestamp value. Once an event piece has been built, this information can be erased from the buffer so that new FE data can be stored. L1 input buffer has been sized through simulation to behave as a time window where all the information related to the same event is stored and no output information from FE is lost due to buffer overload. From the results in fig. \ref{L1_sim} regarding maximum data frame length (integrated charge related data from the FE with the same timestamp) and maximum number of event pieces in the L1, the buffer size required is smaller than 50 kbits. 

\subsection{Data Compression}
\par
\begin{figure}[!t]
	\centering
	\includegraphics[width=3.25in]{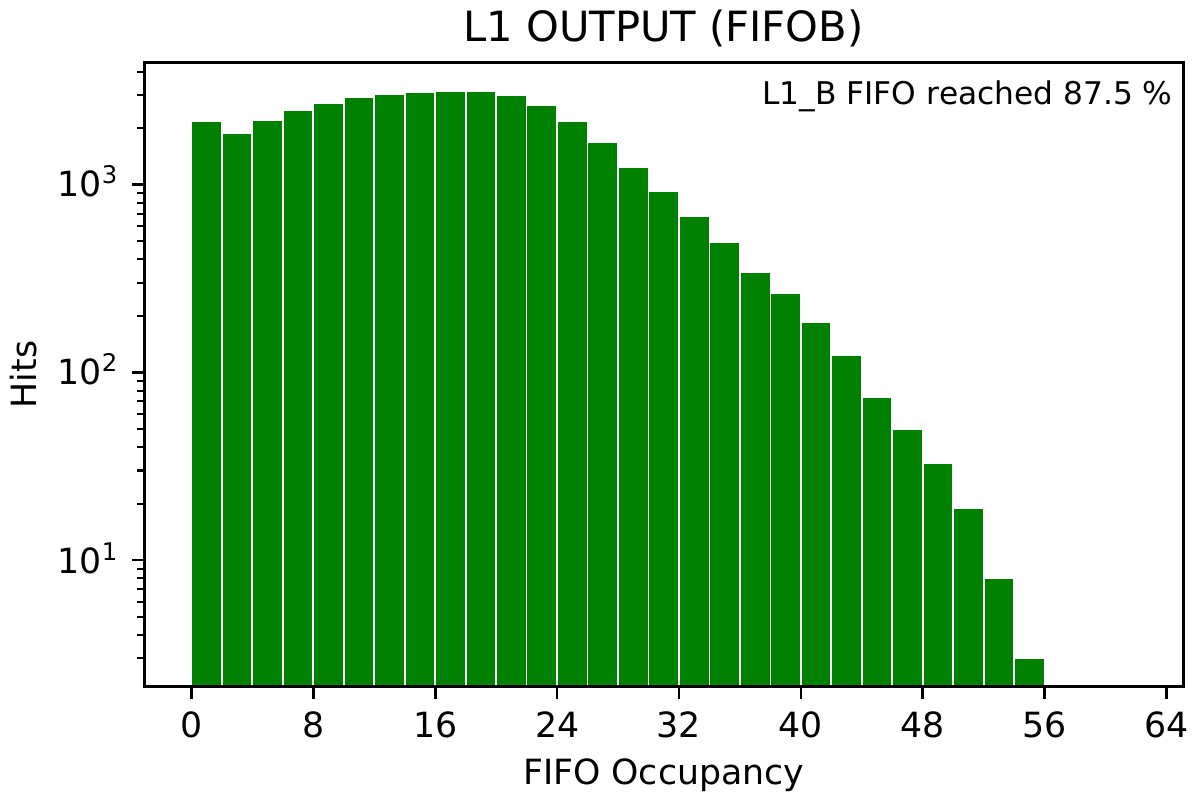}
	% where an .eps filename suffix will be assumed under latex, 
	% and a .pdf suffix will be assumed for pdflatex; or what has been declared
	% via \DeclareGraphicsExtensions.
	\caption{L1 FIFO size simulation.}
	\label{L1_FIFO_size}
\end{figure}

 L1 output data link must operate at a speed of 2.2 Gbits/s for the required simulation conditions and its output FIFO must store a maximum of 60 data frames (54 kbits) (fig. \ref{L1_FIFO_size}). Although this size is affordable for any FPGA device, the output data link speed is quite high. Moreover the amount of data generated for 12 L1 would reach 25 Gbits/s for the storage system at the end of the readout. This link speed could be optimized if a compressor element is introduced in L1.     

\par
\begin{figure}[!t]
	\centering
	\includegraphics[width=3.25in]{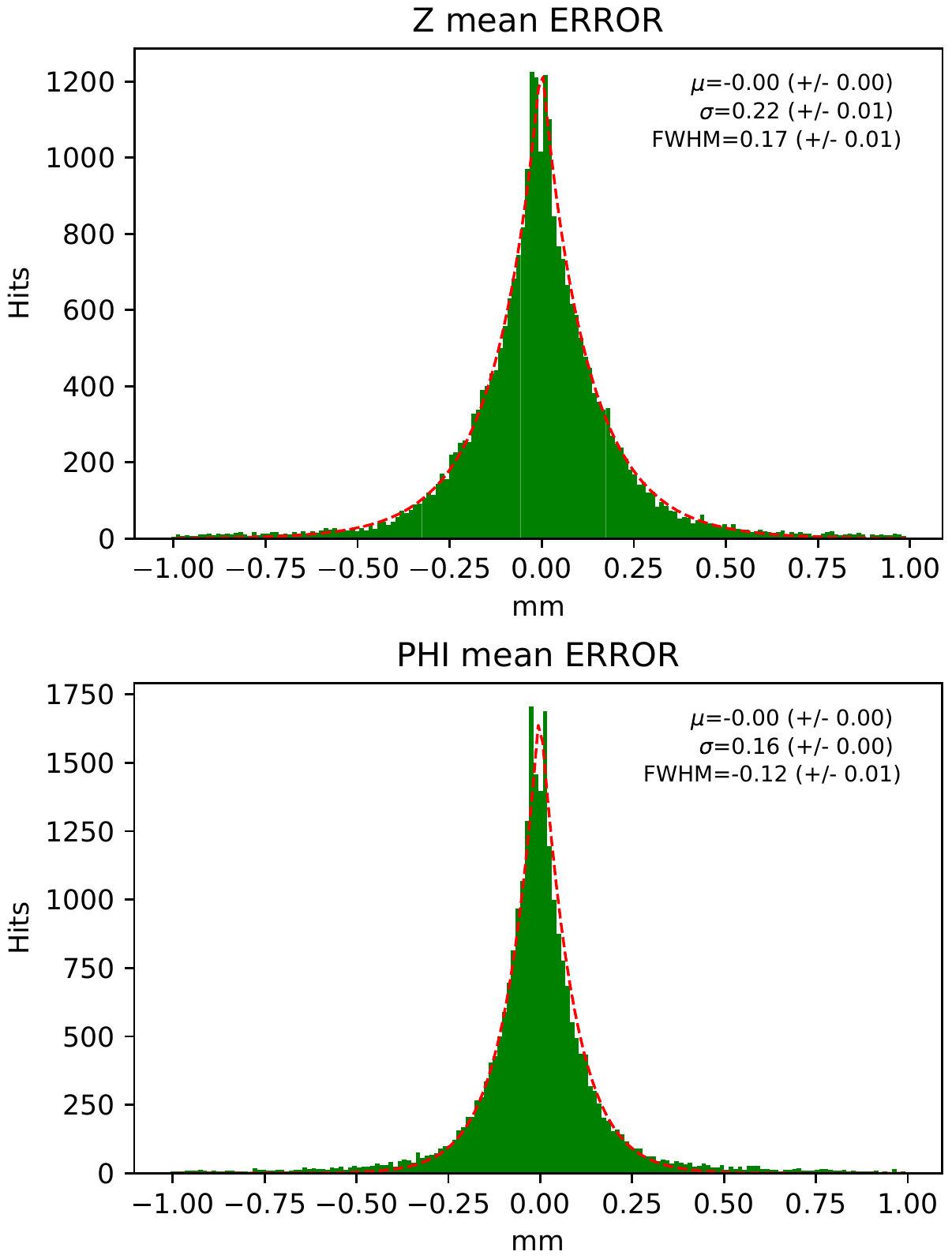}
	% where an .eps filename suffix will be assumed under latex, 
	% and a .pdf suffix will be assumed for pdflatex; or what has been declared
	% via \DeclareGraphicsExtensions.
	\caption{Resolution loss due to compressor.}
	\label{wavelet_res}
\end{figure}

 As explained in previous section, each data frame generated at the output of L1 has a single timestamp and belongs to the same event. Since that data frame composes a piece of the same gamma event, any image based compression techniques could be applied to reduce data at the output of L1. A single stage Haar-1 wavelet filter has been evaluated. This type of compressor extracts low spatial frequency and high spatial frequency components of the original image and uses a lower number of bits to quantize the later ones (10 bits / 4 bits) thus reducing the whole amount of data with a minimum loss in image quality. Simulations have been carried out to quantify the effect on spatial resolution of the detected events (fig. \ref{wavelet_res}) and the results show a degradation in Z and PHI coordinates lower than 0.2 mm (FWHM). 

\par
\begin{figure}[!t]
	\centering
	\includegraphics[width=3.25in]{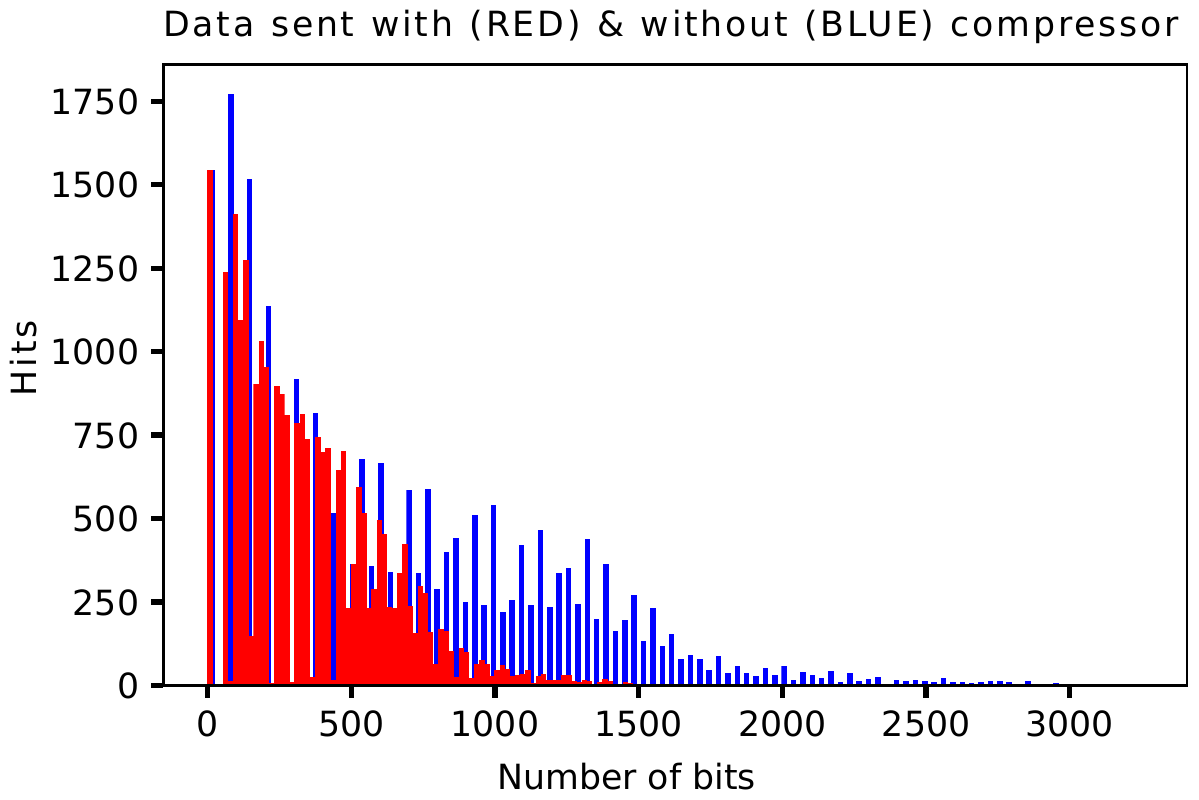}
	% where an .eps filename suffix will be assumed under latex, 
	% and a .pdf suffix will be assumed for pdflatex; or what has been declared
	% via \DeclareGraphicsExtensions.
	\caption{Compressor effect on data frame length}
	\label{compression}
\end{figure}

 The compression rate achieved with this simple filter is around 55\%. However its performance profile (fig. \ref{compression}) shows a higher compression factor for events which involve a higher number of SiPM channels, that is to say those that generate longer data frames and require more time to be transmitted. This fact has a strong impact on data link speed that can be reduced down to 480 Mbits/s for every L1 without data loss.

\section{Conclusions}

\par A new readout concept has been introduced which is compatible with a fully continuous medium detector such as PETALO. The possible drawbacks associated with the amount of data generated as well as the required data link speeds along the system have been analyzed using a custom developed simulation test bench. Results show that the system is feasible however a fast compression technique has been also introduced in L1 to further reduce the data bandwidth requirements.

\ifCLASSOPTIONcaptionsoff
  \newpage
\fi

\bibliographystyle{IEEEtran}
\bibliography{IEEEabrv,NSS_MIC_viherbos}

\end{document}